\documentclass[aps,prl,10pt,twocolumn,amsmath,amssymb,superscriptaddress,preprintnumbers]{revtex4-2}
\usepackage{graphicx}
\usepackage{bm}
\usepackage{hyperref}
\newcommand{\gsim}{\;\rlap{\lower 3.5 pt \hbox{$\mathchar \sim$}} \raise 1pt \hbox {$>$}\;}
\newcommand{\lsim}{\;\rlap{\lower 3.5 pt \hbox{$\mathchar \sim$}} \raise 1pt \hbox {$<$}\;}
\begin{document}

\preprint{\vbox{\hbox{TTP26-033        \hspace{-6mm}}
                \hbox{P3H-26-066       \hspace{-6mm}}				
                \hbox{ZU-TH 32/26      \hspace{-6mm}}				
                \hbox{TTK-26-30        \hspace{-6mm}}				
                \hbox{SI-HEP-2026-19   \hspace{-6mm}}				
                \hbox{CERN-TH-2026-206 \hspace{-2cm}}
}}

\title{\boldmath The Inclusive $\bar B \to X_s \gamma$ Decay Rate with Higher Precision}

\author{M.~Misiak}
\affiliation{Institute of Theoretical Physics, Faculty of Physics, University of Warsaw, PL-02-093 Warsaw, Poland}
\author{H.~M.~Asatrian}
\affiliation{Yerevan Physics Institute, 0036 Yerevan, Armenia}
\author{K.~Brune}
\affiliation{PRISMA++ Cluster of Excellence \& Mainz Institute for Theoretical Physics, Johannes Gutenberg University, D-55128 Mainz, Germany}
\author{M.~Czaja}
\affiliation{Institut f\"ur Theoretische Teilchenphysik, Karlsruhe Institute of Technology, D-76131 Karlsruhe, Germany}
\author{M.~Czakon}
\affiliation{Institut f\"ur Theoretische Teilchenphysik und Kosmologie, RWTH Aachen University, D-52056 Aachen, Germany}
\author{M.~Fael}
\affiliation{Dipartamento di Fisica e Astronomia ``Galileo Galilei'', Universit\`a di Padova, IT-35131 Padova, Italy}
\affiliation{Istituto Nazionale di Fisica Nucleare, Sezione di Padova, IT-35131 Padova, Italy}
\author{C.~Greub}
\affiliation{Albert Einstein Center for Fundamental Physics, Institute for Theoretical Physics, University of Bern, CH-3012 Bern, Switzerland}
\author{T.~Huber}
\affiliation{Theoretische Physik 1, Center for Particle Physics Siegen, Universit\"at Siegen, D-57068 Siegen, Germany}
\author{F.~Lange}
\affiliation{Physik-Institut, Universit\"at Z\"urich, CH-8057 Z\"urich, Switzerland}
\affiliation{PSI Center for Neutron and Muon Sciences, CH-5232 Villigen PSI, Switzerland}
\author{L.~T.~Moos}
\affiliation{Theoretische Physik 1, Center for Particle Physics Siegen, Universit\"at Siegen, D-57068 Siegen, Germany}
\author{M.~Niggetiedt}
\affiliation{Physik Institut, Universit\"at Z\"urich, CH-8057 Z\"urich, Switzerland}
\author{A.~Rehman}
\affiliation{Department of Physics, University of Alberta, Edmonton, AB T6G 2J1, Canada}
\author{K.~Sch\"onwald}
\affiliation{CERN, Theory Department, CH-1211 Geneva 23, Switzerland} 
\author{M.~Steinhauser}
\affiliation{Institut f\"ur Theoretische Teilchenphysik, Karlsruhe Institute of Technology, D-76131 Karlsruhe, Germany}

\begin{abstract}

We present a new and upgraded analysis of the inclusive weak radiative
decay of the $B$ meson in the standard model and beyond. The
${\mathcal O}(\alpha_s)$ perturbative corrections are included in a
formally complete manner. At the order $\alpha_s^2$, exact
dependence on the charm quark mass is calculated for the dominant
corrections, which removes a sizeable uncertainty that was present
in all the previous analyses. The global normalization factor and
many of the non-perturbative contributions are expressed in terms of
the kinetic-scheme mass of the $b$-quark and the heavy-quark-expansion
matrix elements. These parameters are adopted from the most recent
semileptonic fits that include ${\mathcal O}(\alpha_s^3)$ corrections
and up-to-date experimental results. We find ${\mathcal B}_{s\gamma}^{\rm
SM} = (3.54 \pm 0.14)\times 10^{-4}$ for the CP- and isospin-averaged
branching ratio of the considered decay in the standard model, with a
lower cut on the photon energy $E_\gamma > 1.6\,{\rm GeV}$. It agrees
with the current experimental average ${\mathcal B}_{s\gamma}^{\rm
exp} = (3.49 \pm 0.19)\times 10^{-4}$, which provides constraints on
beyond standard model physics. In particular, we find $M_{H^\pm} >
670\,{\rm GeV}$ at $95\%\,{\rm C.L.}$ for the charged Higgs boson mass
in the two-Higgs-doublet model~II.

\end{abstract}

\maketitle

\section{I. Introduction \label{sec:intro}}

Weak radiative decays of the $B$ mesons are known to provide important
constraints on popular extensions of the standard model (SM). In the
present Letter, we consider ${\mathcal B}_{s \gamma}$, namely the CP-
and isospin-averaged branching ratio of the inclusive weak radiative
decays of $B^\pm$, $B^0$ and $\bar B^0$ into charmless $|S|=1$ final
states. The current Particle Data Group (PDG) average of the experimental
results~\cite{Chen:2001fja,Aubert:2007my,Lees:2012wg,Lees:2012ym,Belle:2009nth,Saito:2014das}
for this quantity amounts to~\cite{ParticleDataGroup:2026mpi}
\begin{equation} \label{brexp} 
{\mathcal B}_{s \gamma}^{\rm exp} = (3.49 \pm 0.19)\times 10^{-4},
\end{equation}
with a lower cut of $1.6\,{\rm GeV}$ on the photon energy in the
$B$-meson rest frame.

Precise theoretical calculations of ${\mathcal B}_{s \gamma}$ are
possible thanks to the heavy quark expansion (HQE) that allows to
express the dominant contributions to the hadronic inclusive decay
rate $\Gamma(\bar B \to X_s \gamma)$~\cite{footnote1} in terms of the
perturbative $b$-quark decay rate $\Gamma(b \to X_s^p
\gamma)$~\cite{footnote2}, up to corrections that are suppressed by
powers of $\bar\Lambda/m_b$, with $\bar\Lambda \sim m_B-m_b$. To
achieve a similar accuracy as in Eq.~(\ref{brexp}), the perturbative
rate $\Gamma(b \to X_s^p \gamma)$ must be evaluated including strong
interaction corrections of the orders $\alpha_s$ and
$\alpha_s^2$. As far as the non-perturbative ${\mathcal
O}(\bar\Lambda/m_b)$ corrections are concerned, many of them can be
parameterized in terms of the HQE matrix elements that are determined,
in particular, from the measured spectral moments of the inclusive
semileptonic $B$-meson decays~\cite{Manohar:2000dt}.

In the current Letter, we present results of a new and upgraded
analysis of ${\mathcal B}_{s \gamma}$ in the SM.  The ${\mathcal
O}(\alpha_s)$ perturbative corrections are now formally complete,
thanks to including the four- and five-body final state contributions
from Ref.~\cite{Brune:2025zhd}. For some of the important ${\mathcal
O}(\alpha_s^2)$ corrections, their exact dependence on the charm-quark
mass $m_c$ is now determined~\cite{Czaja:2026xxx}, contrary
to the previous calculations~\cite{Misiak:2015xwa,Czakon:2015exa}
where an interpolation between the $m_c=0$ and $m_c \gg m_b$ limits
was applied. Estimates of the non-perturbative effects include
constraints that follow from the measured isospin
asymmetry~\cite{Belle:2018iff}, as well as from the updated analysis
of the so-called resolved photon contributions in
Ref.~\cite{Gunawardana:2019gep}. Moreover, the HQE matrix elements are
adopted from the most recent kinetic-scheme
fit~\cite{Carvunis:2025vab} that includes ${\mathcal O}(\alpha_s^3)$
corrections~\cite{Fael:2020njb,Fael:2020tow}.

The Letter is organized as follows: In the next section, calculations
of the new perturbative contributions are summarized. A brief
discussion of the non-perturbative effects is presented in
Sec.~III.
 Our main result for ${\mathcal B}_{s \gamma}$ in
the SM is given in Sec.~IV, 
together with sample constraints on beyond-SM physics.
We conclude in Sec.~V.

\section{II. Perturbative corrections} \label{sec:pert}

Calculations of ${\mathcal B}_{s \gamma}$ are conveniently performed
in the low-energy effective theory (LEFT) framework, after decoupling
of the $W$ boson and all the heavier particles at the electroweak
scale $\mu_0 \sim M_W, m_t$. Renormalization group evolution of the
LEFT Wilson coefficients down to the scale $\mu_b \sim m_b/2$ resums
large QCD logarithms $\alpha_s \log(\mu_0/\mu_b)$. The dominant
contributions to $\Gamma(\bar B \to X_s \gamma)$ originate from the
following four LEFT operators:
\begin{eqnarray} 
Q_1  &=& (\bar{s}_L \gamma_{\mu} T^a c_L) (\bar{c}_L     \gamma^{\mu} T^a b_L)~ ,\nonumber\\
Q_2  &=& (\bar{s}_L \gamma_{\mu}     c_L) (\bar{c}_L     \gamma^{\mu}     b_L)~ ,\nonumber\\
Q_7  &=&  \frac{e}{16\pi^2}   m_b (\bar{s}_L \sigma^{\mu \nu}     b_R) F_{\mu \nu}~  ,\nonumber\\
Q_8  &=&  \frac{g_s}{16\pi^2} m_b (\bar{s}_L \sigma^{\mu \nu} T^a b_R) G^a_{\mu \nu}~ . \label{operators}
\end{eqnarray}
Let $\hat G_{ij}^{(2)}$ parameterize the ${\mathcal O}(\alpha_s^2)$
contributions to $\Gamma(b \to X_s^p \gamma)$ that originate from
interferences of amplitudes with $Q_i$ and $Q_j$ interactions. In the
SM, the matrix $\hat G_{ij}^{(2)}$ is real symmetric, up to negligible
${\mathcal O}(V_{ub}/V_{cb})$ corrections. The elements $\hat
G_{77}^{(2)}$ and $\hat G_{78}^{(2)}$ are known in a complete manner
since a long time~\cite{Melnikov:2005bx,Blokland:2005uk,Asatrian:2006ph,Asatrian:2006sm,Asatrian:2006rq,Asatrian:2010rq}.
As far as $\hat G_{ij}^{(2)}$ with $i,j \in \{1,2,8\}$ are concerned,
contributions to them from the two-body $b \to s\gamma$ decay channel
are also known, while the three- and four-body channel contributions have so
far been calculated~\cite{Ligeti:1999ea,Ferroglia:2010xe,Misiak:2010tk}
only in the Brodsky-Lepage-Mackenzie (BLM) approximation~\cite{Brodsky:1982gc}.

\begin{figure}[t]
\begin{center}
\includegraphics[width=38mm,angle=0]{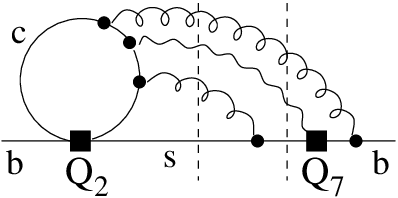}
\caption{\sf A sample diagram contributing to $\hat{G}^{(2)}_{27}$,
with two possible unitarity cuts marked by the vertical dashed lines.
Solid, curly and wavy lines represent quarks, gluons and
  photons, respectively.
\label{fig:diags}}
\end{center}
\end{figure}

Numerically, the most important new effect in our present analysis
originates from $\hat G_{17}^{(2)}$ and $\hat G_{27}^{(2)}$. The
corresponding interferences of amplitudes can be expressed in terms of
four-loop propagator diagrams with unitarity cuts -- see
Fig.~\ref{fig:diags}. All such diagrams with arbitrary $m_c$ and $m_b$
were calculated in Ref.~\cite{Czaja:2026xxx} to which we refer the
reader for technical details and the final renormalized
results. Earlier, unrenormalized results for the two-body
contributions alone were found in
Refs.~\cite{Greub:2023msv,Fael:2023gau,Czaja:2023ren,Greub:2024mwp}.

When compared to the previous determination~\cite{Misiak:2015xwa,Czakon:2015exa}
of $\hat G_{17}^{(2)}$ and $\hat G_{27}^{(2)}$ where an interpolation
between the $m_c=0$ and $m_c \gg m_b$ limits was used, the new
exact-$m_c$ calculation~\cite{Czaja:2026xxx} leads to an enhancement
of ${\mathcal B}_{s \gamma}$ by around $4.2\%$, which exceeds the
previously estimated $\pm 3\%$ ``one-sigma'' interpolation uncertainty
by a factor of $1.4$.  However, the overall ${\mathcal O}(\alpha_s^2)$
correction to ${\mathcal B}_{s \gamma}$ for the central values of our
parameters and renormalization scales is not large, only around
$2.5\%$ -- see Fig.~\ref{fig:mudep} in Sec.~IV.

It should be noted that no cut on the photon energy was imposed in the
exact-$m_c$ calculation~\cite{Czaja:2026xxx} of $\hat G_{17}^{(2)}$
and $\hat G_{27}^{(2)}$, similarly to the interpolated
case~\cite{Misiak:2015xwa,Czakon:2015exa}. Such an additional
approximation is applied only in these particular two contributions to
${\mathcal B}_{s \gamma}$. Removing this approximation in the future
is expected to have a minor effect, similarly to what is observed at
${\mathcal O}(\alpha_s)$ in the same interference terms. The
corresponding photon spectra are peaked in the vicinity of the
endpoint $E_\gamma \simeq m_b/2$, and only narrow tails extend below
$1.6\,{\rm GeV}$. The necessary calculation would involve the same
diagrams with three- and four-body unitarity cuts (see
Fig.~\ref{fig:diags}) but with an additional constraint on $E_\gamma$
imposed using the method of Ref.~\cite{Melnikov:2005bx}. The
constraint introduces an additional scale to the four-loop Feynman
integrals, which makes their calculation considerably more involved.

Apart from the ${\mathcal O}(\alpha_s^2)$ terms, our current analysis
includes new four- and five-body ${\mathcal O}(\alpha_s)$
contributions from Ref.~\cite{Brune:2025zhd} that originate from the
four-quark LEFT operators with $(\bar sb)(\bar q q)$ flavor content,
where $q \in \{u,d,s\}$. Their numerical effect on ${\mathcal B}_{s
\gamma}$ is small -- it amounts to around $-0.2\%$ at our central
point. However, only after their inclusion, the ${\mathcal
O}(\alpha_s)$ calculation of $\Gamma(b \to X_s^p \gamma)$ can be
called formally complete.

\section{III. Non-perturbative effects} \label{sec:nonp}

Some of the perturbative contributions to $\Gamma(b \to X_s^p
\gamma)$, including the ones mentioned at the end of the previous
section, contain collinear logarithms $\log(m_b/m_q)$, which signals
the presence of non-perturbative effects that are not suppressed by
$\bar\Lambda/m_b$. They can be estimated using fragmentation
functions, as in the analyses of
Refs.~\cite{Ferroglia:2010xe,Kapustin:1995fk,Asatrian:2013raa}.  It
turns out that the fragmentation function estimates can roughly be
reproduced by varying $\log(m_b/m_q)$ in the purely perturbative
expressions within the range $[\log 10, \log 50] \simeq [
\log(m_B/m_K), \log(m_B/m_\pi) ]$. Such a rough approach is sufficient
thanks to extra suppression factors that the considered corrections
come with: phase-space factors due to $E_\gamma > 1.6\,{\rm GeV}$,
small Wilson coefficients or the Cabibbo-Kobayashi-Maskawa ratio
$V_{ub}/V_{cb}$ in the case of $(\bar sb)(\bar q q)$ operators, as
well as the squared down-quark charge $Q_d^2 = 1/9$ in the case of
$Q_8$-$Q_8$ interference. For the latter interference, we adopt the
uncertainty estimate from Ref.~\cite{Benzke:2010js} that gives around
$\pm 0.9\%$ in ${\mathcal B}_{s \gamma}$ in the vicinity of our
central point. It is meant to include both the collinear and 
${\mathcal O}(\bar\Lambda/m_b)$ effects. A more recent re-analysis of
this contribution in Ref.~\cite{Hurth:2023paz} resolves certain
theoretical issues but contains no update of the numerical uncertainty
estimate. As far as the remaining collinear terms are concerned, their
effect on ${\mathcal B}_{s \gamma}$ changes by less than $0.4\%$
when $\log(m_b/m_q)$ is varied in the $[\log 10, \log 50]$ range.

Non-perturbative corrections to the $Q_7$-$Q_7$ interference start at
${\mathcal O}(\bar\Lambda^2/m_b^2)$ and amount to around $-4\%$ in
${\mathcal B}_{s \gamma}$. They have been determined up to ${\mathcal
O}(\bar\Lambda^3/m_b^3, \alpha_s \bar\Lambda^2/m_b^2)$ -- see
Ref.~\cite{Ewerth:2009yr} and references therein. They are
parameterized in terms of the HQE matrix elements that we adopt
from the semileptonic fit of Ref.~\cite{Carvunis:2025vab},
along with $m_b$, $|V_{cb}|$, and the corresponding
correlation matrix. Including the correlations is essential, otherwise
a considerable overestimate of the parametric uncertainty would be
obtained.

The $Q_7$-$Q_8$ and $Q_7$-$Q_{1,2}$ interferences receive
non-perturbative ${\mathcal O}(\bar\Lambda/m_b)$ corrections from the
so-called resolved photons in the $Q_8$ and $Q_{1,2}$ amplitudes. In
these amplitudes, scattering of the partonic $b$-quark decay products
with the $B$-meson remnants produces energetic photons at distances of
the order $1/\bar\Lambda$ from the $b$-quark
annihilation vertex. In the $Q_7$-$Q_8$ case, one can take advantage
of the fact that the effect is related to the measured isospin
asymmetry, and reduce the corresponding uncertainty to around $\pm
0.7\%$ -- see section~3 of Ref.~\cite{Misiak:2020vlo} for details. In
the $Q_7$-$Q_{1,2}$ case, we follow the Soft-Collinear Effective
Theory analyses of Refs.~\cite{Benzke:2010js,Gunawardana:2019gep}. The
resulting effect on ${\mathcal B}_{s \gamma}$ is around $+3\%$ -- see
appendix~B of Ref.~\cite{Czaja:2026xxx} for details.

Non-perturbative effects matter also for the photon energy cut choice
$E_\gamma > E_0 = 1.6\,{\rm GeV}$. The local HQE in the dominant
$Q_7$-$Q_7$ interference is applicable when $m_b (m_b - 2 E_0) \gg
\bar\Lambda^2$. If $E_0$ is too close to $m_b/2$, the integrated decay
rate and spectral moments depend on the $B$-meson shape functions --
see the SIMBA Collaboration analysis in
Ref.~\cite{Bernlochner:2020jlt} and references therein. The plots in
that paper confirm that $E_0 = 1.6\,{\rm GeV}$ is well within the
local HQE region. The PDG average of ${\mathcal B}_{s \gamma}$ in
Eq.~(\ref{brexp}) has been obtained after extrapolating in $E_0$ from
the actually applied experimental cuts in the $[1.8,2.0]\,{\rm GeV}$
range down to $E_0 = 1.6\,{\rm GeV}$. However, the PDG extrapolation
factors and their uncertainties should still be updated, which could
most conveniently be done using the very SIMBA results from
Ref.~\cite{Bernlochner:2020jlt}. On the other hand, the SIMBA approach
of performing the theory-experiment comparison using the photon energy
spectrum properties rather than ${\mathcal B}_{s \gamma}$ itself needs
to be updated to include the new perturbative results from the current
Letter. Otherwise their uncertainties are understandably about twice
larger than ours, as explained in Ref.~\cite{Czaja:2026xxx}. Such an
update on the SIMBA side should be straightforward once $\hat
G_{17}^{(2)}$ and $\hat G_{27}^{(2)}$ are determined in the future
without neglecting the photon energy cut.

\section{IV. Final results} \label{sec:res}

Our final result for ${\mathcal B}_{s \gamma}$ in the SM reads
\begin{equation} \label{brsm}
{\mathcal B}_{s\gamma}^{\rm SM} = \left( 3.54 \pm 0.14 \right) \times 10^{-4},
\end{equation}
at $E_0 = 1.6\,{\rm GeV}$. The overall uncertainty of around $\pm 4\%$
has been obtained by combining in quadrature the $\pm 3\%$
higher-order uncertainty and the parametric uncertainty of around $\pm
2.7\%$. A complete list of our input parameters can be found in
appendix~C of Ref.~\cite{Czaja:2026xxx}. It includes, in particular,
several factors whose ranges are fixed by estimates of the
non-perturbative resolved photon contributions. As far as the
higher-order uncertainty is concerned, it is assumed to cover all the
approximations at the ${\mathcal O}(\alpha_s^2)$ level, our truncation
of the perturbative series at this level, as well as the unknown
higher-order non-perturbative contributions. The latter case refers
both to higher orders in $\bar\Lambda/m_b$ and in $\alpha_s$.
\begin{figure}[t]
\begin{center}
\includegraphics[width=7cm,angle=0]{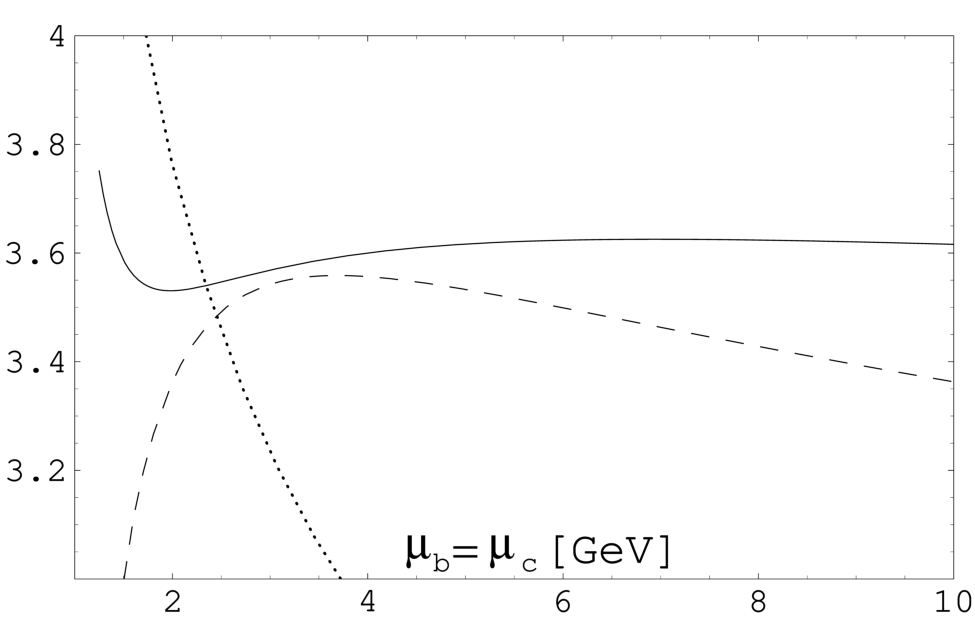}
\caption{\sf ${\mathcal B}_{s \gamma}\times 10^4$ at the leading order
(dotted curve), ${\mathcal O}(\alpha_s)$ (dashed curve), and
${\mathcal O}(\alpha_s^2)$ (solid curve) as a function of the
renormalization scale $\mu_b$, in the case $\mu_b=\mu_c$ (see the text).
The electroweak matching scale $\mu_0$ is fixed at $160\,{\rm GeV}$.
\label{fig:mudep}}
\end{center}
\end{figure}

As usual, lower bounds on the higher-order uncertainties can be
derived by studying the renormalization scale dependence. The plot in
Fig.~\ref{fig:mudep} shows the dependence of ${\mathcal B}_{s \gamma}$
on the renormalization scale $\mu_b$ at subsequent orders of the
perturbative expansion in $\alpha_s$. Non-perturbative corrections are
included in all the three curves in the same manner. The charm quark
mass is $\overline{\rm MS}$-renormalized at the scale $\mu_c$ that is
set equal to $\mu_b$ in the presented plot. It is clearly visible that
inclusion of the ${\mathcal O}(\alpha_s^2)$ corrections considerably
reduces the renormalization scale dependence. The values of
${\mathcal B}_{s \gamma}$ described by the solid line change by around
$\pm 1.3\%$ when the low-energy scale ($\mu_b=\mu_c$) is varied in the
range $[1.5,10]\,{\rm GeV}$. An analogous reduction of the scale
dependence is even more pronounced in the case of the electroweak
matching scale $\mu_0$ -- see Fig.~4 of Ref.~\cite{Czaja:2026xxx}.

The SM result in Eq.~(\ref{brsm}) is in perfect agreement with the
experimental one in Eq.~(\ref{brexp}). Although no signal for
beyond-SM physics is found in ${\mathcal B}_{s \gamma}$, we can still
derive powerful constraints on extensions of the SM. A particularly
simple example is the two-Higgs-doublet model (2HDM) that differs from
the SM only by addition of an extra electroweak doublet of spin-0
bosons. Its physical particle spectrum contains, apart from the SM
particles, a single charged scalar $H^\pm$, a single neutral one
$H^0$, and a pseudoscalar $A^0$. The model has several versions
depending on what structure of the Yukawa couplings is assumed. In
the so-called 2HDM-II, the up-type quarks couple to the Higgs doublet
called $H_u$, while the down-type quarks and leptons couple to the one
called $H_d$, exactly as in the case of the minimal supersymmetric
standard model. The third-generation quark mass hierarchy is then
naturally obtained for a large ratio of the Higgs-doublet vacuum
expectation values $\tan\beta \equiv \langle H_u \rangle/\langle H_d
\rangle$. Beyond-SM contributions to ${\mathcal B}_{s \gamma}$ depend
in the 2HDM-II (at the leading order in electroweak interactions) only
on $\tan\beta$ and the charged-scalar mass $M_{H^\pm}$.

\begin{figure}[t]
\begin{center}
\includegraphics[width=7cm,angle=0]{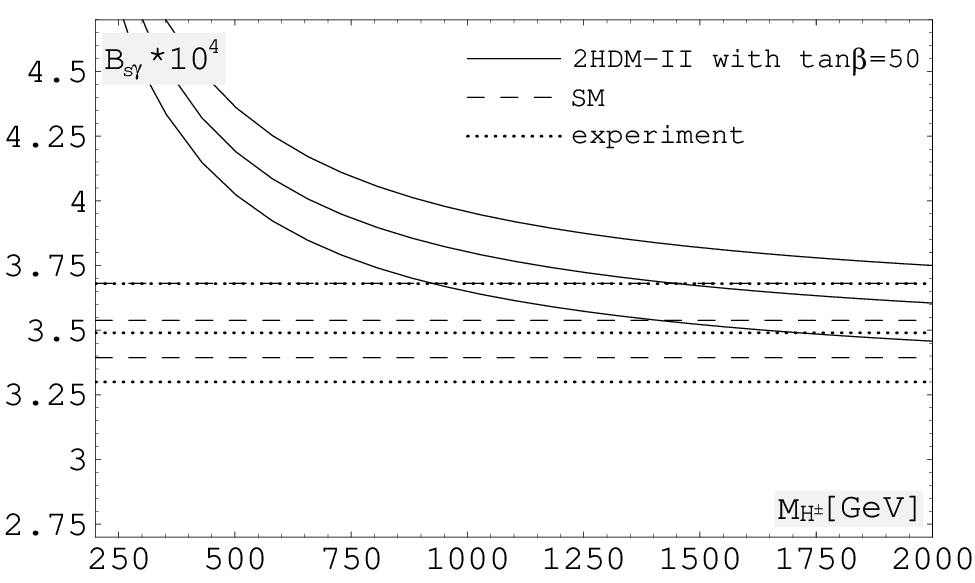}
\caption{\sf ${\mathcal B}_{s \gamma}\times 10^4$ in the 2HDM-II as a
function of the charged scalar mass $M_{H^\pm}$ for fixed $\tan\beta =
50$ (solid lines). The middle line corresponds to the central value,
while the remaining ones indicate the $1\sigma$ range. The dashed and
dotted lines describe the SM result (Eq.~(\ref{brsm})) and the experimental
average (Eq.~(\ref{brexp})) in the same manner. Note that the upper dotted
line is covered by the corresponding dashed one.
\label{fig:MHc2}}
\end{center}
\end{figure}

The solid lines in Fig.~\ref{fig:MHc2} present the dependence of
${\mathcal B}_{s \gamma}$ on $M_{H^\pm}$ in the 2HDM-II for fixed
$\tan\beta = 50$. The middle line is the central value, while the
upper and lower ones indicate the $1\sigma$ uncertainty. The dashed
and dotted lines describe the SM result (Eq.~(\ref{brsm})) and the
experimental average (Eq.~(\ref{brexp})) in the same
manner. Irrespectively of $\tan\beta$, the charged scalar effects in
the 2HDM-II can only enhance ${\mathcal B}_{s \gamma}$, and the
enhancement factor decreases with $\tan\beta$ at fixed
$M_{H^\pm}$. The $\tan\beta = 50$ case is practically equivalent to
the $\tan\beta \to +\infty$ limit. Consequently, the bound on
$M_{H^\pm}$ in the 2HDM-II that can be derived from
Fig.~\ref{fig:MHc2} is an absolute one, i.e.\ it holds for any
$\tan\beta$. Following the approach of Ref.~\cite{Misiak:2017bgg}, we
find $M_{H^\pm} > 670\,{\rm GeV}$ at $95\%\,{\rm C.L.}\,$. No absolute
bound of comparable strength is available in the considered model from
the LHC direct searches.

In other extensions of the SM, bounds from ${\mathcal B}_{s \gamma}$
on their parameter spaces are often complicated, and impossible to
present in such a simple manner as in the 2HDM-II case. The main
beyond-SM contributions that affect ${\mathcal B}_{s \gamma}$ are
usually encoded in modifications of the Wilson coefficients $C_7$ and
$C_8$ at the electroweak matching scale $\mu_0 = 160\,{\rm GeV}$. In
practice, it is sufficient to calculate such modifications (denoted by
$\Delta C_7$ and $\Delta C_8$ below) at ${\mathcal O}(\alpha_s^0)$
only because the new physics effects cannot be large, given the
precise agreement of the SM prediction with the experimental world
average. Using such an approach, we obtain
\begin{equation} 
{\mathcal B}_{s\gamma}\times 10^4 = (3.54 \pm 0.14) - 8.51\,\Delta C_7 - 2.16\,\Delta C_8,
\end{equation}
where the quadratic terms in $\Delta C_{7,8}$ have been neglected. For
more refined analyses of constraints on beyond-SM physics, the above
formula should be extended to include other LEFT operators and/or
${\mathcal O}(\alpha_s)$ corrections to $C_{7,8}(\mu_0)$ that are
available in the generic case from Ref.~\cite{Bobeth:1999ww}. In the
2HDM, also the ${\mathcal O}(\alpha_s^2)$ corrections to $C_{7,8}(\mu_0)$
are known~\cite{Hermann:2012fc} and included in our 2HDM-II
considerations above.

\section{V. Conclusions} \label{sec:concl}

We performed a new and upgraded analysis of ${\mathcal B}_{s\gamma}$
in the SM and beyond. We significantly improved the precision by
evaluating $\hat G_{17}^{(2)}$ and $\hat G_{27}^{(2)}$ for arbitrary
values of $m_c$ and $m_b$, which required performing a four-loop
calculation. In effect, a sizeable uncertainty due to the former
interpolation in $m_c$ is removed. We find an enhancement of
${\mathcal B}_{s\gamma}$ by around $4.2\%$ with respect to the case with
interpolation. In addition, the ${\mathcal O}(\alpha_s)$ calculation
is rendered formally complete by evaluating the formerly missing four-
and five-body final state contributions. The input parameters are
updated to their most recent values, and improvements in estimating
the non-perturbative contributions are taken into account. Our final
SM result is
${\mathcal B}_{s\gamma}^{\rm SM} = \left( 3.54 \pm 0.14 \right) \times 10^{-4}$,
for $E_\gamma > 1.6\,{\rm GeV}$. It is in perfect agreement with the
experimental world average
${\mathcal B}_{s\gamma}^{\rm exp} = (3.49 \pm 0.19)\times 10^{-4}$.
We provide a simple phenomenological formula for ${\mathcal B}_{s\gamma}$
in generic extensions of the SM. In the case of the 2HDM-II, we find an
absolute bound $M_{H^\pm} > 670\,{\rm GeV}$ at $95\%\,{\rm C.L.}\,$.

\begin{acknowledgments} 
We acknowledge partial support from the 
Deutsche Forschungsgemeinschaft (DFG) under grants 396021762 (TRR 257),                          
533766364 (EXC 3107),                                                                                
and 390831469 (EXC 2118/2),                                                                       
from the National Science Center (Poland) under grants 2024/55/B/ST2/01703 and 2023/49/B/ST2/00856,  
from the Higher Education and Science Committee of Armenia Program Grant 21AG-1C084,          
from the European Research Council under grant 101019620 (ERC Advanced Grant TOPUP),                 
from the European Union under the Marie Sk{\l}odowska-Curie Actions (MSCA) grant 101204018,          
and from the Swiss National Science Foundation (SNSF) under
contract~\href{https://data.snf.ch/grants/grant/211209}{TMSGI2\_211209}.
\end{acknowledgments}


\begin{thebibliography}{99}
\bibitem{footnote2} The partonic final states are denoted by $X_s^p$ to avoid confusion with the hadronic states $X_s$.
\end{thebibliography}
\end{document}